\documentclass[aps,pra,a4paper,twocolumn,superscriptaddress,floatfix]{revtex4}

\usepackage{amssymb}
\usepackage{graphicx}
\usepackage{amsmath}
\usepackage{bbm}

\usepackage[pdftex]{hyperref}
\usepackage{color}
\usepackage{natbib}
\usepackage{ulem}
\makeatletter
	\renewcommand{\maketag@@@}[1]{\hbox{\m@th\normalsize\normalfont#1}}
\makeatother

\begin{document}

\title{Quantum optomechanical piston engines powered by heat}
\author{A. Mari}
\author{A. Farace}
\author{V. Giovannetti}
\affiliation{NEST, Scuola Normale Superiore and Istituto Nanoscienze-CNR, I-56127 Pisa}

\begin{abstract}
We study two different models of optomechanical systems where a temperature gradient between two radiation baths is exploited for inducing self-sustained coherent oscillations of a mechanical resonator. From a thermodynamic perspective, such systems represent quantum instances of self-contained thermal machines converting 
heat into a periodic mechanical motion and thus they can be interpreted as nano-scale analogues of macroscopic piston engines. 
Our models are potentially suitable for testing fundamental aspects of quantum thermodynamics in the laboratory and for applications in energy efficient nanotechnology. 
\end{abstract}

\maketitle

\section{Introduction}

The research on quantum thermodynamics received large attention since the beginning of quantum physics. Its main task is understanding to what extent the laws of thermodynamics are valid in the quantum regime \cite{Beretta, Mahler1, Mahler2, QTNano, Oppenheim,renner,jarzynski}. A particularly relevant question is how much can thermal machines (heat engines and refrigerators) be miniaturized while retaining their essential feature of producing work or extracting heat \cite{QHE,popescu1,eisert,adesso,Benenti, Kurizki, campisi}. 

In this paper we propose simple models of quantum piston engines based on optomechanical systems \cite{OM1,OM2}, {\it i.e.}\ devices
composed of micro/nano-scale mechanical resonators coupled to optical or microwave modes. In the last few years  exceptional levels of quantum control over optomechanical systems have been reached. For example important milestones like ground state  cooling of a mechanical resonator \cite{gs,crystal,microwave},  squeezing \cite{squeezing} and optomechanical entanglement \cite{entanglement} have been experimentally achieved.
These facts suggest that the research level on optomechanics is sufficiently advanced to allow implementations of quantum thermodynamics ideas with near-future technology. 

\begin{figure}[t]
\includegraphics[width=0.8 \columnwidth]{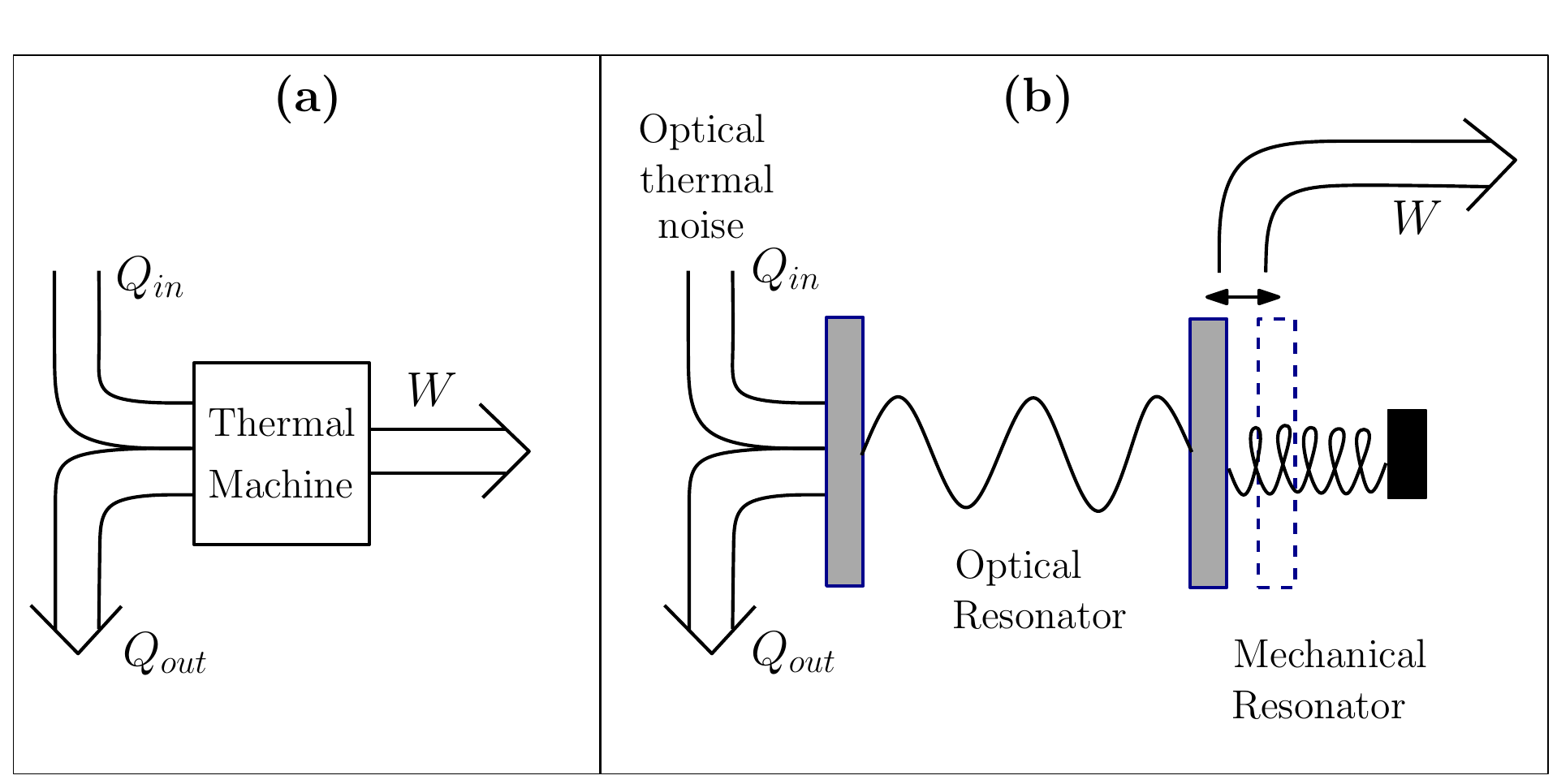}
\caption{ { (a)} Scheme of a general heat engine: input heat $Q_{in}$ is absorbed, output heat $Q_{out}$ is dissipated and output work $W$ is produced.
{ (b)} Model of an equivalent quantum optomechanical thermal motor. Heat is absorbed from a hot optical/microwave thermal bath. This energy is partially used to 
excite the coherent motion of a mechanical resonator (work) and the rest is dissipated into a cold optical/microwave bath. }
 \label{fig1}
\end{figure}

Recently some models of optomechanical engines have been proposed, where the systems are driven by periodic coherent lasers and thermodynamic cycles are induced by a cyclic tuning of the parameters \cite{Elouard,meystre,Hou-Ian}.   For example, in the specific system considered in Ref.\cite{meystre}, a thermodynamic Otto cycle is induced by the modulation the laser detuning. Here instead we propose two self-contained optomechanical setups, that we call {\it single cavity engine} and {\it cascade engine}, in which a temperature gradient between two thermal baths is exploited for inducing self-sustained oscillations (phonon lasing \cite{lasing0, lasing1, lasing2, lasing3, lasing-exp1, lasing-exp2}) of a mechanical resonator, in the absence of external forces and external control. In this sense our approach is similar to the analysis of the finite dimensional thermal machines introduced in \cite{popescu1, quadridot}, to the ``cooling by heating'' setup proposed in \cite{cbh} and to the concept of Brownian motors reviewed in \cite{BM1, BM2}. The emergence of persistent mechanical oscillations  in a system which is subject to friction and dissipation can be interpreted as a continuous production of thermodynamic work.  Indeed, the thermodynamic interpretation of a lasing system as a quantum heat engine can be traced back to the seminal work by Scovil and Schulz-DuBois \cite{maser-engine}  and has been studied more recently in the context of hybrid (continous-descrete) systems \cite{maser-engine2,quantum-optical-engine, Kurizki, photocell}.

The idea of extracting coherent motion from random Brownian noise is actually well known since the famous ratchet machine introduced by  Smoluchowsky and 
Feynman \cite{ratchet}. This phenomenon has been later developed and applied in several different contexts mostly described by classical statistical mechanics \cite{shimizu, BM1, BM2}. In this work we
propose an implementation of a Brownian motor with optomechanical systems and, most importantly, we treat the dynamics taking into account quantum effects. This allows us to focus on the differences between quantum and classical thermal machines and to shed some light on the limitations that the miniaturization of nano-mechanical devices will encounter due to the emergence of quantum effects. An important contribution of this work is also the concept of {\it maximum power under load}. This quantity is well known in the field of mechanical engineering as a figure of merit of macroscopic motors but, up to our knowledge, was never applied to quantum mechanical engines.

After completing this work another model of autonomous optomechanical engine has been proposed \cite{kurizki2}. However, in the model considered in Ref.\ \cite{kurizki2} the extractable work is examined during the initial amplification of the mechanical oscillations and therefore it strongly depends on the initial state of the mechanical resonator, which acts as an external non-equilibrium resource. Differently, the performance of our engines is analyzed in the steady-state and does not depend on the initial conditions.

\section{Optomechanical piston engines}
\begin{figure}[th]
\includegraphics[width=0.9 \columnwidth]{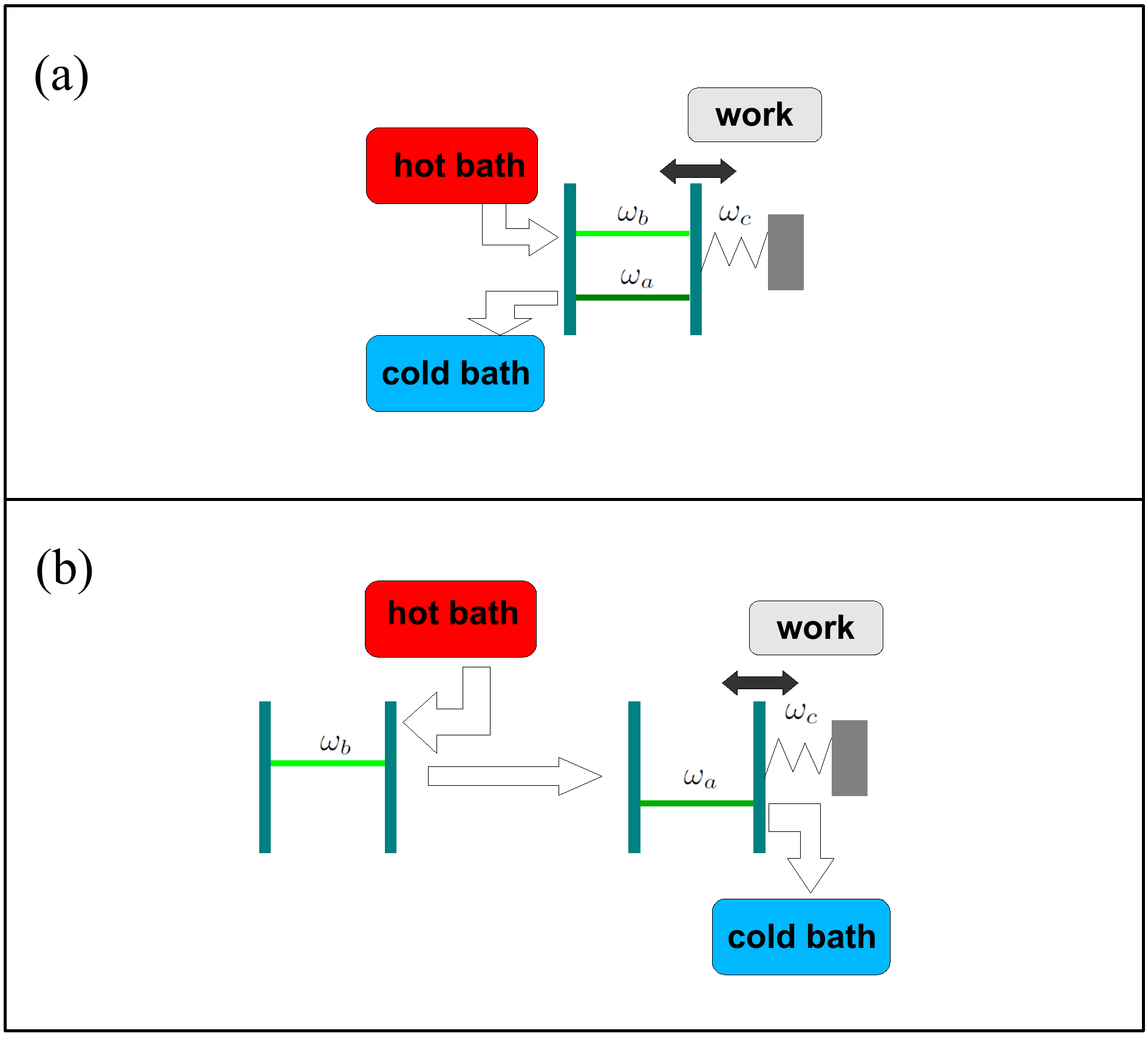}
\caption{ Schemes of the two quantum piston engines considered in this work: (a) single cavity engine, (b) cascade engine. Setups (a) and (b) are formally described by the respective quantum master equations  \eqref{singlecavity} and \eqref{cascade}. }
 \label{FIGURES}
\end{figure}

In this section we introduce our models of optomechanical piston engines. In particular we propose two different quantum engines arranged in a single cavity and in a cascade setup respectively (see Fig.\ \ref{FIGURES}). Eventually we also introduce a completely classical model of the single cavity engine that we are going to compare later with the quantum counterpart. 

\subsection{Single cavity engine}

The first system that we consider involves a mechanical resonator of frequency $\omega_c$ coupled by radiation pressure to two radiation modes
of frequency  $\omega_a$ and $\omega_b$ respectively. The corresponding Hamiltonian is 
\begin{eqnarray}
H=&\hbar \omega_a a^\dag a+\hbar \omega_b b^\dag b + \hbar \omega_c c^\dag c \nonumber \\
    &- \hbar g (a+b)^\dag (a+b) (c+c^\dag),\label{H}
\end{eqnarray}
where $a,b,c$ are the bosonic annihilation operators of the three modes and $g$ is the optomechanical coupling constant. 
The last term in Eq.\ \eqref{H} is proportional to the position of the mechanical resonator and to the intensity of the cavity 
field. This Hamiltonian could describe the radiation pressure of two cavity fields on a moving mirror \cite{OM1,OM2}, but also different systems  like toroidal micro-cavities \cite{toroid}, opto-mechanical crystals \cite{crystal}, cold atoms \cite{BEC}, {\it etc.}. This model applies as well to electro-mechanical systems where the radiation modes have frequencies in the microwave range \cite{microwave,entanglement}.

The three modes are put in contact with three independent environments, which can possess different temperatures.  The corresponding dynamics of the open system, in the weak coupling limit, is well described by the following master equation \cite{Gardiner}: 
\begin{eqnarray}\label{singlecavity}
\dot \rho = - \frac{i}{\hbar}[H,\rho] &+ \kappa_a (N_a+1) D_a( \rho) + \kappa_a N_a D_{a^\dag} (\rho) \nonumber \\
                  &+ \kappa_b (N_b+1) D_b( \rho) + \kappa_b N_b D_{b^\dag} (\rho)  \nonumber \\
                  &+ \kappa_c (N_c+1) D_c( \rho) + \kappa_c N_c D_{c^\dag} (\rho) ,
\end{eqnarray}
where the $D_x(\cdot)$ is the Lindblad dissipator $D_x (\rho)= x \rho x^\dag - \frac{1}{2}\{x^\dag x, \rho\} $ associated with the modes $x=a,b,c$, \ $\kappa_x$ is the decay rate,  and $N_x$ depends on the temperature $T_x$
of the respective environment according to the Bose-Einstein statistics
$ N_x = [e^{\frac{\hbar \omega_x}{k_B T_x}}-1]^{-1} $.  We remark that the assumption of independent heat baths implicit in Eq.\ \eqref{singlecavity} is consistent only if the two modes $a$ and $b$ are spectrally well defined and distinguishable $|\omega_b - \omega_a| < (\kappa_a + \kappa_b)/2$. In particular this implies that the treatment of the single cavity engine is formally valid only for optomechanical systems in the resolved sideband regime $\omega_c< ( \kappa_a+\kappa_b)/2$.

We stress that, compared with standard optomechanical systems, in Eq.\ \eqref{H} there is not a driving laser contribution and the dynamics is induced only by the heat fluxes associated with the hot and cold baths. For what concerns the physical implementation of the hot bath in the laboratory, one could think of a simple black-body light source successively filtered around the frequency $\omega_b$, or alternatively one could use a laser of frequency $\omega_b$ driven below threshold (incoherent regime). The noise bandwidth should be larger than $\kappa_b$ but smaller than the free spectral range $|\omega_b-\omega_a|$, in order not to affect the other mode $a$. The cold bath is instead automatically implemented without any driving by the natural coupling of mode $a$ with the vacuum field outside the cavity.   

As we are going to show, if the resonance condition $\omega_b-\omega_a=\omega_c$ is satisfied and if  the thermal noise parameter 
$N_b $ is large enough, then it is possible to excite mechanical self-sustained oscillations of the mode $c$. Before presenting the results in details, let us first introduce also
the second model of optomechanical engine. 

\subsection{Cascade engine}
In the previous model (single cavity engine) two optical modes are supported by the same optomechanical cavity. For technical reasons it may be more practical to realize a cascade engine where the mode $b$ is associated with an independent optical cavity whose output is fed into a standard optomechanical system based on a single optical mode (see Fig.\ \ref{FIGURES}.b). This setting provides results which are qualitatively equivalent to the single cavity setup and, at the same time,
it could be experimentally easier to realize. For example the required tuning of the resonance condition $\omega_b-\omega_a=\omega_c$  should be much simpler if the two modes $a$ and $b$ are supported by two separated devices. Moreover, since the cascade setup does not require a coherent coupling between the first and the second mode, this theoretical model corresponds to the simple experimental scenario in which incoherent colored noise centered around $\omega_b$ is injected into a standard single-mode optomechanical system.

The Hamiltonians associated with the first and second cavities are respectively:
\begin{eqnarray}
H_1 &=&\hbar \omega_b b^\dag b ,   \\
H_2&=&\hbar \omega_a a^\dag a +\hbar \omega_c c^\dag c- \hbar g a^\dag a (c+c^\dag). \quad
\end{eqnarray}
In addition to the dissipative channels that we introduced in the single cavity engine, here we also have to consider that the output of the first cavity is fed into the second one. The corresponding master equation can be derived using the quantum optics framework of {\it cascaded quantum systems} \cite{Gardiner,cascade}, obtaining (see {\it e.g.} Eq.\ (12.1.16) of Ref.\ \cite{Gardiner}): 
\begin{eqnarray} \label{cascade}
\dot \rho &=& - \frac{i}{\hbar}[H_1+H_2,\rho] \nonumber \\
                && + \kappa_a (N_a+1) D_a( \rho) + \kappa_a N_a D_{a^\dag} (\rho) \nonumber \\
                &  &+ \kappa_c (N_c+1) D_c( \rho) + \kappa_c N_c D_{c^\dag} (\rho) \nonumber \\
                 & &+\gamma_1 D_b (\rho) +\gamma_2 D_a (\rho) 
                                   -\sqrt{\gamma_1 \gamma_2} ( [a^\dag, b \rho]  +[ \rho b^\dag, a])  \nonumber \\
                 & & +\dfrac{N_b}{2}  \left[ [\sqrt{\gamma_1} b +\sqrt{\gamma_2} a ,\rho] , \sqrt{\gamma_1} b^\dag +\sqrt{\gamma_2} a^\dag \right]  \nonumber \\
                 & & +\dfrac{N_b}{2}   \left[ [\sqrt{\gamma_1} b^\dag +\sqrt{\gamma_2} a^\dag ,\rho] , \sqrt{\gamma_1} b +\sqrt{\gamma_2} a \right] . \quad 
\end{eqnarray}
The first three lines of Eq.\ \eqref{cascade} are analogous to the previous model.  The last three lines instead describe a cascade setup in which the light exiting the first cavity with a rate $\gamma_1$ is fed into the second cavity
with a rate $\gamma_2$. In the following we will set for simplicity $\gamma_1=\gamma_2=\kappa_a$, considering a scenario in which the filter and the optomechanical system consist on two cavities with equal finesse.
 
\begin{figure}[!]
\includegraphics[width=  1.03 \columnwidth]{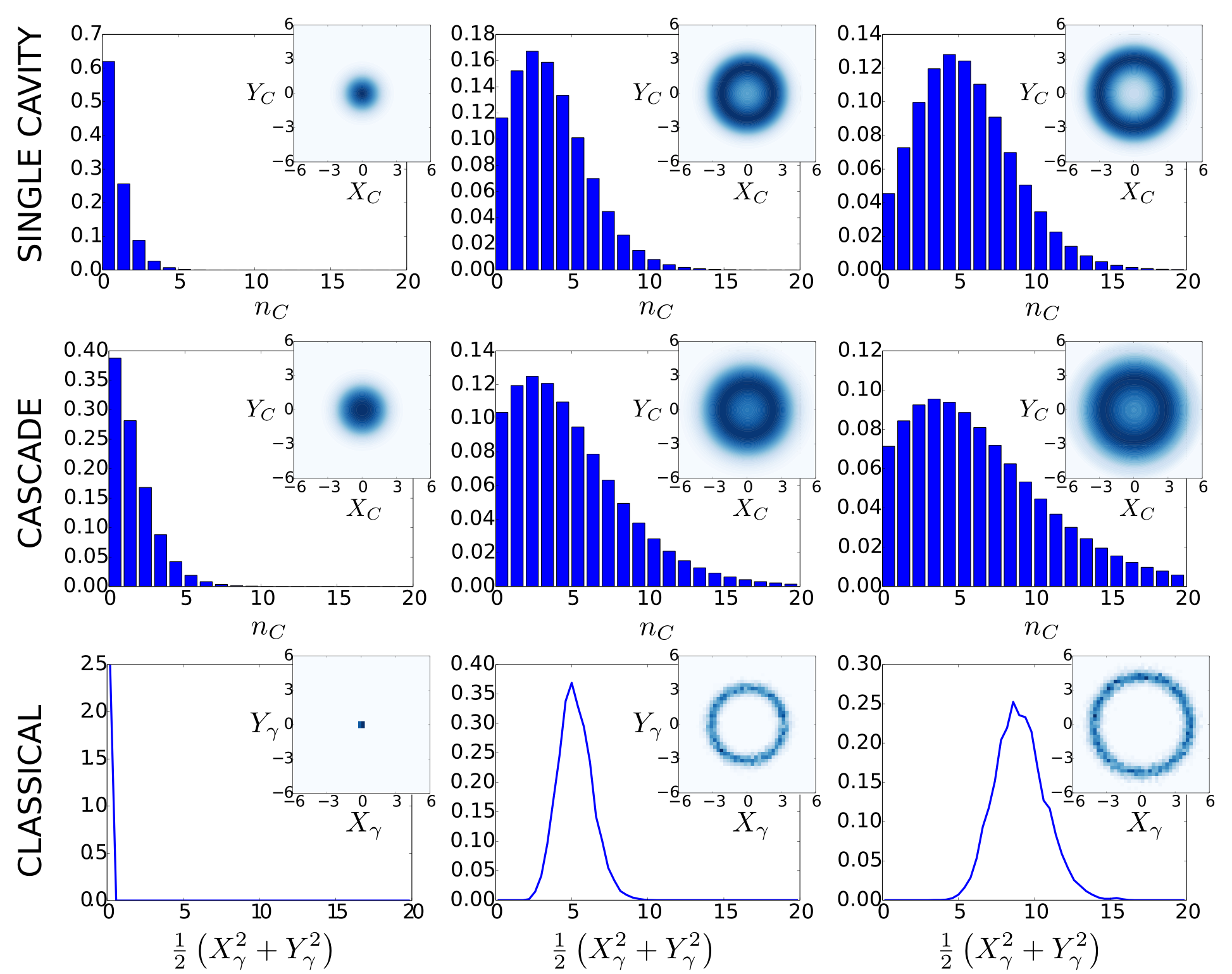} 
\caption{From left to right, phonon number distributions and Wigner functions (insets) for different values of the thermal noise parameter $N_b= 0.17,  0.33,  0.5$ ($N_b=0$ trivially gives the vacuum state). The first row refers to the single cavity engine with parameters $N_a=N_c=0$,
 $\omega_b-\omega_a=\omega_c=1$, $\kappa_a=\kappa_b=0.2$, $\kappa_c=0.005$ and $g=0.06$. The second row refers to the cascade engine with  $\kappa_a=\gamma_1=\gamma_2=0.15$, $\kappa_c=0.003$ and $g=0.1$. The third row instead represents the classical version of the single cavity engine  (see appendices A and B for further details). } \label{limit}
\end{figure}

\subsection{Classical engine} In order to investigate the differences between classical and quantum thermal machines we will also compare the single-cavity engine with its own classical version. The classical model is obtained interpreting the Hamiltonian of Eq.\ \eqref{H} as being described by classical position and momentum quadratures. Replacing the quantum operators $a,b,c$ with classical complex amplitudes $\alpha, \beta, \gamma$ the master equation \eqref{singlecavity} corresponds in the classical model to the following system of Langevin equations 

\begin{align}
\dot{\alpha}&= - i \Delta \alpha  + i g (\alpha + \beta)(\gamma^* + \gamma) - \frac{\kappa_a}{2} \alpha + \xi^a, \nonumber \\
\dot{\beta}&= + i g (\alpha + \beta)(\gamma^* + \gamma) - \frac{\kappa_b}{2} \beta + \xi^b, \nonumber \\
\dot{\gamma}&= - i \omega_c \gamma  + i g |\alpha + \beta|^2 - \frac{\kappa_c}{2} \gamma + \xi^c. \label{stochastic0}
\end{align}
In the above expressions $\kappa_a$, $\kappa_b$ and $\kappa_c$ are the dissipation rates (we keep the same values as in the quantum engine), while $\xi^a=(\xi^a_x+ i \xi^a_y)/\sqrt{2}$, $\xi^b=(\xi^b_x+ i \xi^b_y)/\sqrt{2}$ and $\xi^c=(\xi^c_x+ i \xi^c_y)/\sqrt{2}$ are independent complex zero-mean Gaussian random variables with correlations $\left< \xi^\nu_x(t) \xi^\nu_x(t') \right> = \left< \xi^\nu_y(t) \xi^\nu_y(t') \right> = \kappa_\nu N_\nu \delta(t-t')$ ($\nu=a,b,c $).\\

Differently from the standard theory of optomechanical systems, this model is not a semi-classical approximation of some quantum Langevin equations but represents instead an intrinsically classical description of the optomechanical system, directly obtainable from classical statistical mechanics. Indeed our aim is not to approximate the quantum model, but to understand the differences between the quantum and the classical engines.
More details on the derivation of the classical Langevin equations \eqref{stochastic0} and on the specific methods used for their numerical simulation are given in appendices A and B.

\section{Self-sustained oscillations powered by heat}
The possibility of inducing coherent self-sustained oscillations in optomechanical systems has been theoretically \cite{lasing0,lasing1, lasing2, lasing3} and experimentally demonstrated \cite{lasing-exp1,lasing-exp2}, and it is nowadays a well established technique. However, in all works except \cite{lasing-exp2}, this effect is induced by coherent external drivings.  In Ref. \cite{lasing-exp2}, oscillations of a non-linear mechanical system are excited by electrical colored noise, but only in the classical regime.
In our quantum engines there is not a driving term in the Hamiltonian and the only source of energy is provided by the incoherent absorption of heat. It is therefore not guaranteed that coherent oscillations at the quantum level can emerge in our setups and the main task of this work is to give a proof of principle demonstration that this effect is actually possible. In a second step we will study some thermodynamic aspects of the engines and compare the classical and quantum versions of the motors. 

In the standard theory of optomechanical limit cycles, the driving laser is blue-detuned with respect to the cavity resonance.  From this fact we learn that, if we wish to have self-sustained oscillations in our engines, energy should be put in the radiation mode of larger frequency while the other mode should be as pure as possible in order to absorb and dissipate the photons scattered by the mechanical resonator. The optimal choice of temperatures is therefore $N_a=N_c=0$ and $N_b > 0$.  For the other system parameters we consider typical values which are known to allow limit cycles in the presence of a coherent laser \cite{lasing2,lasing3}  and, as we are going to 
show, these values remain suitable also in our dissipative setups. The specific parameters are reported in the caption of Fig.\ \ref{limit} and are consistent with 
the recent experimental advances in strongly coupled optomechanical systems \cite{crystal, microwave, entanglement, BEC}. We then vary the temperature of the bath of the mode $b$ ({\it i.e.} we increase $N_b$) and we numerically solve the steady state condition $\dot \rho=0$ associated to the master equation of the single cavity engine  \eqref{singlecavity} and of the cascade engine \eqref{cascade}. The steady state is found exactly (without rotating wave approximations) in a truncated Fock space of up to 3 photons for the modes $a$ and $b$ and $20$ phonons for the mode $c$.
 
The numerics has been performed using the toolbox QuTiP2 \cite{qutip2}, and the results 
are shown in Fig.\ \ref{limit}. From the sequence of Wigner functions evaluated for increasing values of $N_b$ it is clear that the mechanical resonator is initially heated up in a thermal state and, above a given threshold, it develops a limit-cycle with the characteristic ring shape in phase space. The same 
effect is evident also in the probability distribution of the number of phonons in  the system  (diagonal elements of $\rho$ in the Fock basis), where the transition is from a Gibbs distribution to a Poissonian one typical of a coherent state.  We can thus claim that,  in this regime our optomechanical engines are effectively behaving as quantum piston engines converting heat into coherent mechanical oscillations. 

\begin{figure}[th]
\includegraphics[width=0.97 \columnwidth]{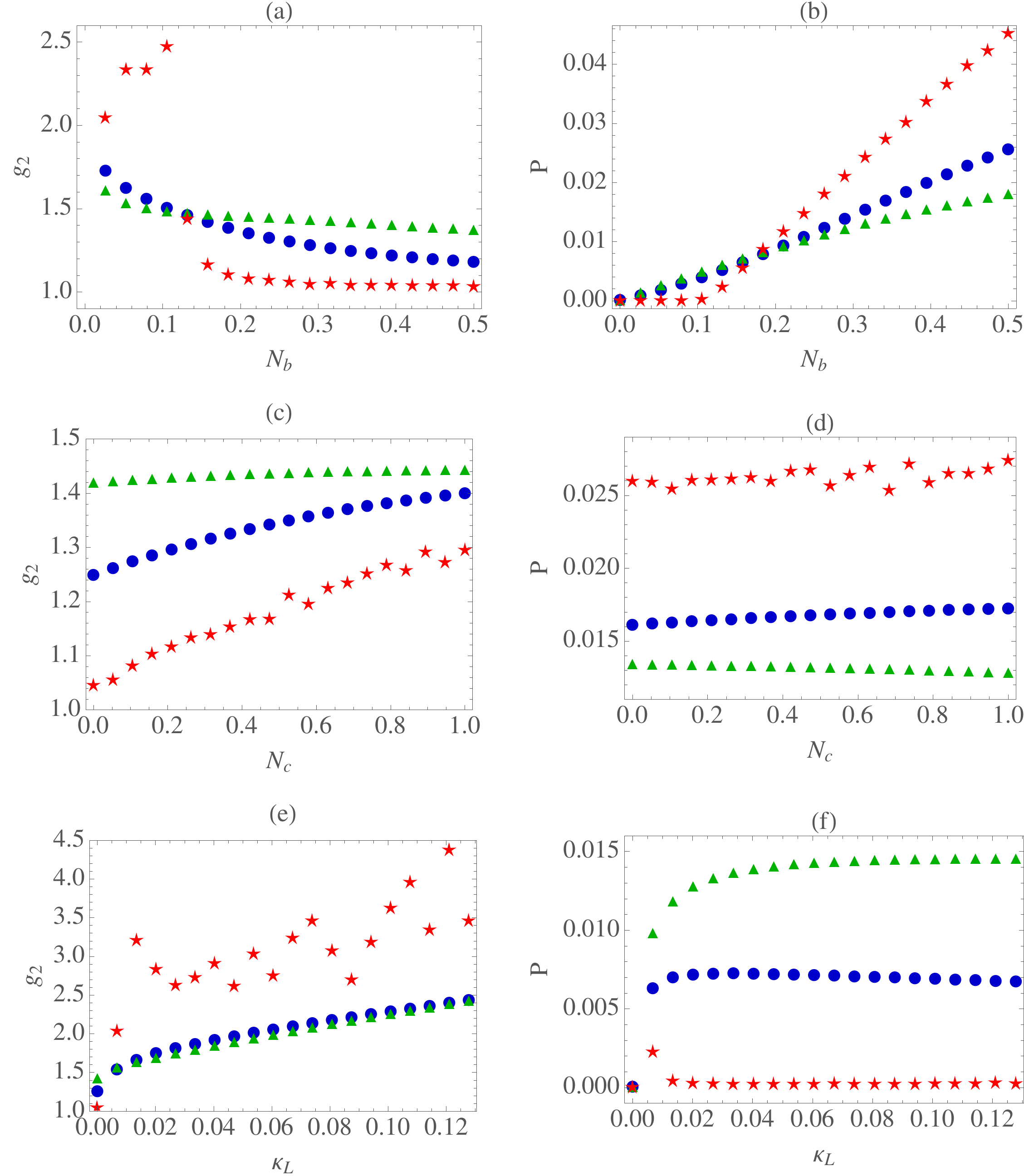}  
\caption{(a) Equal-time second order coherence function $g_2$  and (b) dissipated power $P$ (in units of $\hbar \omega_c$ per second) with respect to the
thermal noise parameter $N_b$. The other parameters are the same as those used in Fig.\ \ref{limit}. 
In (c) and (d), the same quantities are plotted with respect to the mechanical bath mean phonon number $N_c$, for $N_b=0.33$. In (e) and (f), $g_2$ and the 
external power $P_L$ (in units of $\hbar \omega_c $ per second) are evaluated with respect to the load damping rate $\kappa_L$. The maxima of $P_L$ are highlighted by black circles. For all plots the legend is: single cavity engine (blue circles), cascade engine (green triangles), classical engine (red stars).}  \label{fig-g2-power}

\end{figure}

A remark should be made about the notion of ``coherent" oscillations. From the shape of the Wigner function one can see that the steady state of the mechanical oscillator is actually phase randomized and the density matrix is essentially diagonal in the Fock basis. The randomization of the phase is the unavoidable consequence of the rotation symmetry of the system and corresponds exactly to the same feature possessed by the steady states of standard optical lasers. The notion of coherence which then applies in our case is the standard criterion used in quantum optics for distinguishing between thermal and coherent  radiation, namely the equal-time normalized second-order coherence function \cite{Gardiner}: 
$g_2=\langle c^\dag c^\dag  c c \rangle / \langle c^\dag  c \rangle^2 .$
Basically $g_2$ measures how likely it is to consecutively detect two phonons at a given instant of time. For thermal states  one has   $g_2=2$ (bunching statistics), while for coherent states $g_2=1$  (Poissonian statistics).If the quantity $g_2$ decreases from the thermal threshold of $2$ towards lower values, then this is a hint that the field is developing some level of coherence and therefore that a lasing effect is happening in the system.
In Fig.\ \ref{fig-g2-power}.a the quantity $g_2$ is plotted for different values of $N_b$, quantitatively showing the transition 
from an incoherent state to a coherent state of the mechanical resonator. 

The reader may wonder why the curves associated with the classical engine appear more noisy with respect to the quantum counterparts. This is due to the statistical error obtained when averaging the different phase-space trajectories corresponding to independent simulations of the classical dynamics. For the quantum engines instead, apart from the truncation of the Fock space, all the quantities reported in the figures correspond to precise expectation values evaluated on the exact density matrix of the system. 

\section{Performances of the optomechanical engines}

In what follows we study the performances of the previously studied quantum and classical engines.

As noticed in \cite{maser-engine} and further investigated in \cite{maser-engine2,afterburner,quantum-optical-engine,photocell}, the energy of a  coherent field can be interpreted as thermodynamic work and the lasing device as a heat engine. However a quantitative and rigorous analysis of the work produced by the engine is a non-trivial fundamental problem. 
The task of quantifying the maximum work (or power) extractable form a quantum system is still subject to 
a significant research effort \cite{Mahler1, Mahler2, Beretta, campisi, Oppenheim, renner, popescu3,eisert, Kurizki,anders, aberg, janzing}. 

The dynamics of our motors is non-cyclic and open (non-unitary) and it is not straightforward to define the amount of work produced by 
a mechanical resonator which is continuously sustained in a non equilibrium steady state.  We then tackle this problem from  two different directions:
we first adopt a pragmatic approach and try to give an estimate of the power by indirectly considering the flux of energy dissipated by the mechanical resonator into its environment ({\it dissipated internal power}) or, alternatively,  dissipated by an external load connected to the system ({\it external power under load}). 
The second approach is instead based on the framework of the so called {\it resource theories of thermodynamics} \cite{Oppenheim, janzing, renner, aberg, eisert, anders, popescu3, esposito}, under the ideal assumption that after reaching the steady state the mechanical resonator can be detached from the thermal machine and used as a non-equilibrium resource for an arbitrary work extraction task. Clearly the first approach is interesting for its practical interpretation and experimental testability, on the other hand, the second approach has the advantage of providing a much more rigorous estimate of the extractable work at the price of 
assuming an ideal and abstract experimental scenario. 

\subsection{Dissipated internal power} 
Let us start with the pragmatic approach and characterize the performances of our piston engines by estimating the power dissipated by the mechanical resonator.  
The net flux of energy dissipated into the environment can be explicitly computed \cite{maser-engine2} giving :
\begin{eqnarray}\label{P}
P&=&-{\rm Tr}\{\hbar \omega_c c^\dag c \left [ \kappa_c (N_c+1) D_c(\rho)+ \kappa_c N_c D_c^\dag (\rho) \right] \}  \nonumber \\
&=&\hbar \omega \kappa_c ( \langle c^\dag c \rangle_\rho-N_c).
\end{eqnarray}
The  dissipated power has some obvious problems since it does not distinguishes between useless energy (heat) resulting from Brownian fluctuations  and useful energy (work) (a similar issue has been discussed in Ref.\ \cite{quantum-optical-engine}). Nonetheless at least when we are in the lasing regime, {\it i.e.} when the motion of the mechanical mode is coherent, the dynamics is similar to a classical harmonic oscillator rotating in a deterministic circular phase-space orbit. For a classical oscillator it is clear that the energy of the limit cycle can be easily converted into useful work. 
Then we can argue that, if a system is in a coherent limit cycle, the dissipated power is a reasonable figure of merit 
of the work extractable form the system.  It would be an interesting problem to understand how a quantum mechanical limit cycle can be rectified in order to lift a ``weight'' (excite a work medium) in the quantum regime, in the same spirit of \cite{popescu3, Oppenheim, anglin}. This and other quantitative thermodynamic analysis are, however, outside the  ``proof of principle'' approach of this work and will be investigated elsewhere. 

In Fig.\ \ref{fig-g2-power}.b the dissipated power is shown
as a function of $N_b$ for the single-cavity, the cascade, and the classical engines. We observe that the two quantum models are qualitatively equivalent while for large $N_b$ the classical engine seems more powerful than the quantum counterpart. Interestingly however, due to a sharper lasing transition, the classical engine is not able to excite the mechanical resonator for small values of $N_b$. These discrepancies could be associated to the presence of quantum fluctuations in the dynamics of the quantum engine: these fluctuations are deleterious for large $N_b$ but, by smoothing the lasing transition, they became
advantageous for small values of $N_b$. 
Finally in Fig.\ \ref{fig-g2-power}(c) and \ref{fig-g2-power}(d) we report the quantities $g_2$ and $P$ for different temperatures of the mechanical bath. The dissipated power
is slightly modified for larger values of $N_c$, while the quantity $g_2$ is instead increased because the state of the oscillator is inherently partially thermal for $N_c \neq 0$ . A possible interpretation of this thermal regime is the fact is that, for $N_c\neq 0$, a large fraction of power is dissipated in the form of heat rather than work. This problem motivates the necessity of a better figure of merit based on the free energy of the system as explained at the end of this section.

\subsection{External power under load} 
The quantity $P$ introduced in Eq.\ \eqref{P} is the intrinsic power dissipated by the piston due its own mechanical friction. However, in analogy with macroscopic engines, one could imagine to use the mechanical oscillations for doing work on an external load. This load, will act as an additional friction force damping the mechanical mode. Then we can model a generic load substituting $\kappa_c \rightarrow \kappa_c+\kappa_L$ in Eq.s \eqref{singlecavity} and \eqref{cascade}, where $\kappa_L$ is the damping constant due to the load.  The power $P_L$ externally dissipated by the load will be given by  the same expression of Eq.\ \eqref{P} but with the substitution $\kappa_c \rightarrow \kappa_L$:
\begin{eqnarray}\label{PL}
P_L&=&-{\rm Tr}\{\hbar \omega_c c^\dag c \left [ \kappa_L (N_c+1) D_c(\tilde \rho)+ \kappa_L N_c D_c^\dag (\tilde \rho) \right] \}  \nonumber \\
&=&\hbar \omega \kappa_L ( \langle c^\dag c \rangle_{\tilde{\rho}}-N_c),
\end{eqnarray}
where also the steady state $\tilde \rho$ depends on $\kappa_L$ since it is obtained assuming a total mechanical friction of $\tilde \kappa=\kappa_c+\kappa_L$.

It is clear that the external power is zero in the two limits  of $\kappa_L=0$ ({\it i.e.} when there is no load) and of $\kappa_L \rightarrow \infty$ ({\it i.e.} when load is so strong that the thermal machine cannot work). Thus there must exist an optimal load $\kappa_L$ maximizing $P_L$ and it makes sense then to define the {\it maximum power under load}  $P_L^*=\max_{\kappa_L} P_L$ as a possible figure of merit of the performances of a quantum engine when subject to an external load. 

The predicted behavior of $P_L$ as a function of $\kappa_L$ is confirmed by our numerical results presented in Fig.\ \ref{fig-g2-power}.f, where we can observe the existence of an optimal load $\kappa_L^*$ maximizing the external power.  We notice a strong qualitative difference between the classical and the quantum engines. In the quantum case the optimal load is much larger than the intrinsic dissipation rate $\kappa_L^* \gg \kappa_c$ while in the classical case $\kappa_L^*$ is comparable to $\kappa_c$. This difference could be associated to the strong effect of quantum fluctuations in the considered low energy regime.

From the analysis of Fig.\ \ref{fig-g2-power}.e it also evident that the optimal power $P_L^*$ is obtained for a mechanical steady state that is not anymore in the lasing regime because of the strong external dissipation.  This fact suggest that $P_L^*$ is a good figure of merit only if we are interested in the total amount of energy that the optomechanical system can transfer to the load. If instead we are concerned with the more specific task of work extraction, a smaller value of $\kappa_L$ may be more appropriate with respect to $\kappa_L^*$,  because high values of the load decrease the coherence of the mechanical steady-state. We will further comment on this point in section IV D.

\subsection{Non-equilibrium steady state as a resource for work extraction}

Finally we introduce yet another way of characterizing our piston engines based on the maximum work extractable form the non-equilibrium steady state of the mechanical resonator. Ideally one could imagine to detach the mechanical resonator from the thermal machine, and to use it as a ``resource" for some arbitrary work extraction protocol (we neglect the energy cost of the detachment process). This approach, even if quite abstract with respect to the previous estimation of the dissipated power, has the advantage of having a rigorous operational interpretation in terms of resource theories of thermodynamics \cite{Oppenheim, janzing, renner, aberg, eisert, anders, popescu3, esposito}. From such theories we can directly borrow the main results without the need of explaining the details of their derivation. 

A general and fundamental result is the following (see {\it e.g.}\ \cite{Oppenheim,  esposito}): given a system in a quantum state $\rho$ and subject to a given Hamiltonian $H$,  the maximum extractable work (in units of $k_B T$) is upper bounded by the relative entropy between $\rho$ and the Gibbs state $\rho^G_H=e^{-\frac{H}{k_B T}}/\mathcal Z$ with Hamiltonian $H$:

\begin{equation}\label{boundW}
W^{\rm max} \le k_B T  S (\rho \| \rho^G(H) ),
\end{equation}
where $S(\rho_1\| \rho_2)= {\rm Tr } [ \rho_1 \log (\rho_1)-  \rho_1 \log (\rho_2) ] $   (in this work every logarithm is with respect to base $e$). 
By explicitly computing $\log(\rho^G_H)= - H/ (k_B T) - \log(\mathcal Z)$, one can rewrite Eq.\ \eqref{boundW}  in a form analogous to the second law of thermodynamics 
\begin{equation}\label{secondlaw}
W^{\rm max} \le F(\rho) - F(\rho^G_H)=\Delta F,
\end{equation}
where $F(\rho)$ is the non-equilibrium free energy defined as $F(\rho)= {\rm Tr}[\rho H] - k_B T S(\rho)$ and $S(\rho)$ is the Von Neumann entropy of the quantum state.

Moreover, depending on the particular assumptions and on the specific operations allowed by the corresponding resource theory, the bound \eqref{secondlaw} can be saturated with arbitrary good precision. More  
precisely, it can be shown that there exist protocols extracting the maximum work $\Delta F$ in the following cases: in all models based on Hamiltonian quenches composed with complete thermalizations \cite{anders, esposito}, and
also in other models where the work is stored in a given quantum system explicitly included in the theory \cite{Oppenheim, janzing, renner, aberg, eisert, popescu3}. In the latter case however the bound can be saturated only in the thermodynamic limit of infinitely many copies of the resource state $\rho$ \cite{Oppenheim}, or with the use of an additional catalytic coherent system \cite{aberg}. 

\begin{figure}[t] 
\includegraphics[width=0.97 \columnwidth]{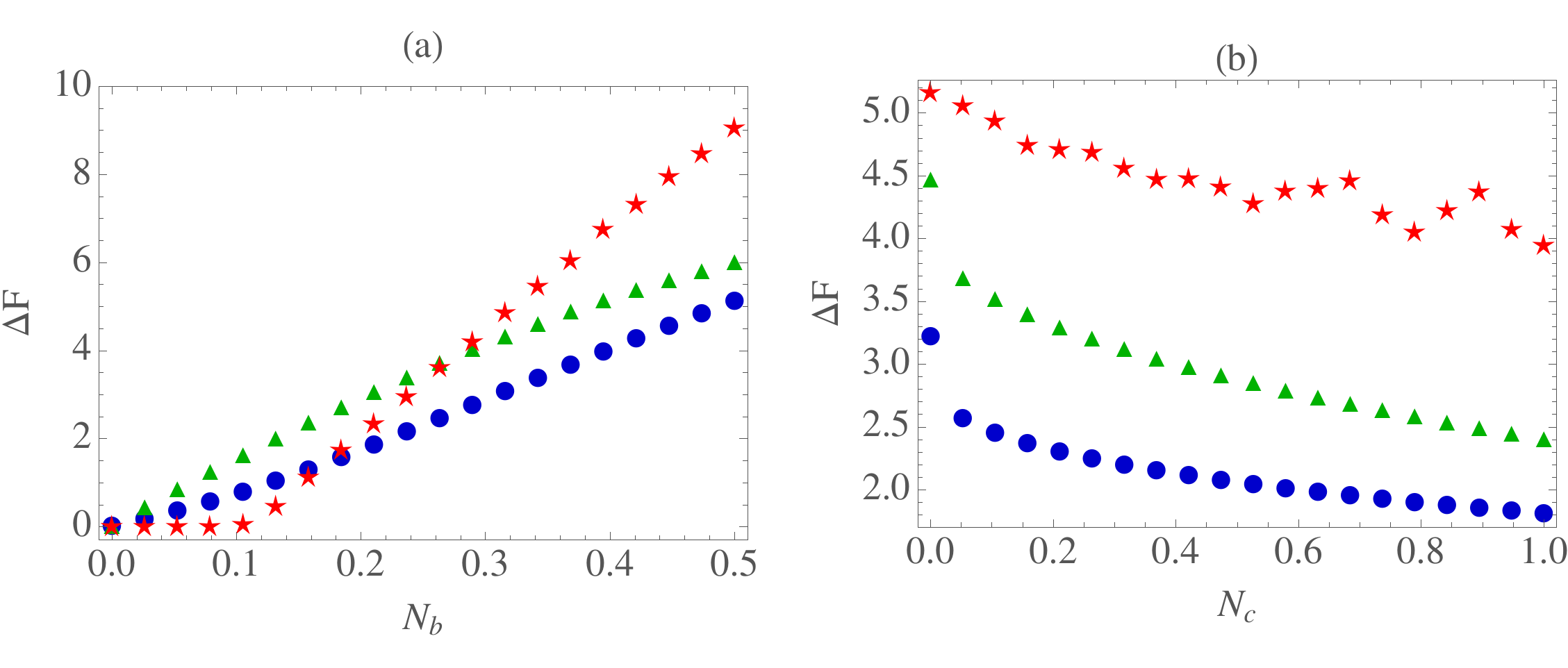}
\caption{(a) Maximum extractable work $\Delta F$ (in units of $\hbar \omega_c$) with respect to the
thermal noise parameter $N_b$. (b) Maximum extractable work $\Delta F$ (in units of $\hbar \omega_c$) with respect to the
thermal noise parameter $N_c$.  For both plots the legend is: single cavity engine (blue circles), cascade engine (green triangles), classical engine (red stars).}  \label{fig-F}
\end{figure}

On the basis of the previous discussion, we use the free energy difference $\Delta F$ appearing in Eq.\ \eqref{secondlaw}, as a figure of merit of the maximum work capability of the optomechanical piston engines considered in the previous section. In Fig. \ref{fig-F}, we plot $\Delta F$ for the steady state of the mechanical resonator of the single cavity engine, the cascade engine and the classical engine. For computing the free energy in the quantum case we take as $H$ the local Hamiltonian of the mechanical mode $\hbar \omega_c c^\dag c$ and as temperature $T$ the one of the mechanical bath (implicitly linked to the parameter $N_c=[e^{\frac{\hbar \omega_c}{k_B T_c}} -1]^{-1}$).  Notice that in Fig. \ref{fig-F}.a the temperature of the mechanical bath is zero and $\Delta F$ reduces to the mean energy of the mechanical mode. On the contrary, in Fig.\ \ref{fig-F}.b the temperature is nonzero and the free energy difference $\Delta F$  is affected by the entropy of the steady state and by the free energy of the Gibbs state $F(\rho^G_H)=k_B T_c \log(1-e^{-\frac{\hbar \omega_c}{k_B T_c}})$.

For computing the free energy in the classical case, we approximate the steady state distribution of the mechanical resonator in the ${ X_{\gamma},Y_{\gamma} }$ phase-space through numerical simulations. Then we extract the quantity ${ F = E - k_{B} T H }$, where $E$ is the mean energy and $H$ is the Shannon entropy of the stationary probability distribution. We also simulate the dynamics of the system in the absence of optomechanical coupling (${ g=0 }$) to extract the free energy of the Gibbs state in an analogous way. More details can be found in appendix B.

In agreement with the analysis of the dissipative power, a direct observation of Fig.\ \ref{fig-F} suggests that the single-cavity engine and the cascade engine are qualitatively equivalent for the task of work extraction, while the classical engine is characterized by a sharp threshold behavior. As expected, the maximum extractable 
work is negatively affected by the thermal noise of the mechanical bath $N_c$.

\subsection{Comparison between different figures of merit}
In this section we discuss the validity and the range of applicability of the three different figures of merit that we have previously introduced (dissipated internal power,  power under load, and free energy difference). 

The {\it dissipated internal power}, defined in Eq.\eqref{P}, measures the net energy flux dissipated by the mechanical resonator into the environment. The motivation behind this quantity is the fact that, for a Gibbs state in equilibrium with the heat bath, there is a balance between absorbed and emitted energy and the net flux is zero. If instead the state is kept out of equilibrium a net amount of energy can constantly flow into the environment. In this stationary non-equilibrium regime, it is clear that the engine must provide at least the amount of power necessary for compensating such dissipated power. 

If an external load is attached to the mechanical resonator, energy can be dissipated not only into the environment but also through the load acting as an additional dissipative channel. In order to sustain the system in the steady state, the engine should provide additional energy which we called {\it power under load} and is given by the expression \eqref{PL}.
Therefore, the operational meaning which we can  associate to the internal power and to the power under load has the nature of an ``energy cost". These quantities are advantageous from an experimental point of view since they can be estimated without the knowledge of the system state and dynamics, but only through a direct measure of the energy flux. However the interpretation of the dissipated energy as  ``work'' or ``heat" is problematic, especially at large  temperatures.

  From classical and quantum thermodynamics we know that energy can be used to perform work only if it is ``ordered", while one should consider ``disordered" energy as heat. The change in free energy is exactly the thermodynamical potential which quantifies the energy of a system which can be converted into useful work in a reversible transformation. For this reason, we used the {\it free energy difference} between the steady state of the mechanical resonator and the Gibbs state as our third figure of merit. This quantity has the operational interpretation of ``extractable work" which is fundamentally different form the ``energy cost" measured by the internal and external dissipated power.  Unfortunately, in order to measure the free energy in the laboratory one needs to, directly or indirectly, measure the entropy of the mechanical resonator and this could be experimentally demanding .

Having in mind the conceptual meanings of the dissipated power and free energy difference, we can compare the corresponding results presented in Fig.\ \ref{fig-g2-power} and Fig.\ \ref{fig-F}.
In particular we observe that Fig.\ \ref{fig-g2-power}.b is qualitatively equivalent to Fig.\ \ref{fig-F}.a apart from a multiplicative factor $\kappa_c$ (which is different for the single cavity and cascade engines).
The equivalence is due to the fact that at zero temperature $N_c=0$ the mean energy and the free energy coincide.  Instead, for $N_c>0$, the dissipated energy plotted in Fig.\ \ref{fig-g2-power}.d has
clearly a different behavior with respect to the free energy difference shown in Fig.\ \ref{fig-F}.b. This discrepancy is due to the fact that the dissipated energy contains both work-like and a heat-like contributions while
the free energy measures only the work-like contribution which is degraded by the temperature of the environment by a quantity $K_B T \Delta S$. 
This can be seen also by comparing Fig. \ref{fig-F}.b with Fig.\ \ref{fig-g2-power}.c. There we observe that when the free energy decreases in Fig. \ref{fig-F}.b, the $g_2$ in Fig. \ref{fig-g2-power}.c increases and approaches the value $2$, meaning that the steady-state is more affected by thermal fluctuations that contribute to heat instead of work.

\section{Conclusions}

We proposed two different models of quantum optomechanical engines based on a single cavity and a cascade setup respectively.  In both cases we have shown that random thermal fluctuations of optical or microwave fields can be exploited for inducing self-sustained coherent oscillations of a mechanical resonator. In this  regime,  
our systems behave as nano-scale analogues of macroscopic piston engines driven by thermal energy. We estimated the dissipated power, the external power under load and the maximum work extractable from the mechanical resonator. We  also highlighted the differences between quantum and classical optomechanical engines. 

We believe that our analysis, together with other recent ideas \cite{Douarche, cbh, Elouard, meystre, kurizki2, brunelli}, could pave the way for the development of fundamental experiments on quantum thermodynamics based on optomechanical systems.
At the same time the paradigm of piston engines presented in this work could find practical technological applications in the fabrication of micro-mechanical motors \cite{BM1,BM2} and energy efficient nano-scale devices \cite{photocell, Gammaitoni,Wang}. 

\noindent {\it Acknowledgments --}
The authors are grateful to R. Fazio, M. Campisi and A. Tomadin for discussions. This work is partially supported by the EU Collaborative Project TherMiQ (grant agreement 618074) and 
by the ERC through the Advanced Grant n. 321122 SouLMan.

\appendix
\section{Classical piston engine}

In this appendix we describe the classical analogue of the quantum single cavity engine presented in the main text. The single cavity engine involves a mechanical resonator of frequency $\omega_c$ coupled by radiation pressure to two radiation modes
of frequency  $\omega_a$ and $\omega_b$ respectively. For the quantum case we have the Hamiltonian (1) given in the main text which gives the following Heisenberg equations for the bosonic annihilation operators $ a,  b,  c$ of the three modes:
\begin{align}
\dot{ a}&= - i \Delta  a  + i g ( a +  b)( c^\dagger +  c), \nonumber \\
\dot{ b}&= + i g ( a +  b)( c^\dagger +  c), \nonumber \\
\dot{ c}&= - i \omega_c  c  + i g ( a +  b)^\dagger( a +  b) . \label{Heisen}
\end{align}
Please note that we expressed the radiation operators in a frame rotating with frequency $\omega_b$. We also defined the detuning $\Delta=\omega_a - \omega_b$ and set it to be $\Delta=-\omega_c$.\\

We obtain the classical counterpart of Eq.s\ \eqref{Heisen} by demoting the operators $ a,  b,  c$ to classical dynamical complex amplitudes $\alpha,\beta,\gamma$ (essentially reversing the standard quantization procedure):
\begin{align}
\dot{\alpha}&= - i \Delta \alpha  + i g (\alpha + \beta)(\gamma^* + \gamma), \nonumber \\
\dot{\beta}&= + i g (\alpha + \beta)(\gamma^* + \gamma), \nonumber \\
\dot{\gamma}&= - i \omega_c \gamma  + i g  | \alpha + \beta|^2 . \label{classic}
\end{align}
The three oscillators are put in contact with three independent environments, which can possess different temperatures. We then add friction terms and classical Brownian noises to Eq.s\ \eqref{classic}, turning them into classical Langevin equations:
\begin{align}
\dot{\alpha}&= - i \Delta \alpha  + i g (\alpha + \beta)(\gamma^* + \gamma) - \frac{\kappa_a}{2} \alpha + \xi^a, \nonumber \\
\dot{\beta}&= + i g (\alpha + \beta)(\gamma^* + \gamma) - \frac{\kappa_b}{2} \beta + \xi^b, \nonumber \\
\dot{\gamma}&= - i \omega_c \gamma  + i g |\alpha + \beta|^2 - \frac{\kappa_c}{2} \gamma + \xi^c. \label{stochastic}
\end{align}
In the above expressions $\kappa_a$, $\kappa_b$ and $\kappa_c$ are the dissipation rates (we keep the same values as in the quantum engine), while $\xi^a=(\xi^a_x+ i \xi^a_y)/\sqrt{2}$, $\xi^b=(\xi^b_x+ i \xi^b_y)/\sqrt{2}$ and $\xi^c=(\xi^c_x+ i \xi^c_y)/\sqrt{2}$ are independent complex zero-mean Gaussian white noises with correlations $\left< \xi^\nu_x(t) \xi^\nu_x(t') \right> = \left< \xi^\nu_y(t) \xi^\nu_y(t') \right> = \kappa_\nu N_\nu \delta(t-t')$ ($\nu=a,b,c $).\\

We stress that this model should not be interpreted as a semi-classical approximation of the quantum system. Instead it represents a purely classical description of the optomechanical system which, in principle, one could derive directly form classical electromagnetism. Indeed our aim is not to approximate the quantum model with the classical one but, on the contrary, to understand the differences between the quantum and the classical optomechanical engines. 

\section{Numerical simulation of the classical engine}

\begin{figure}[t]
\includegraphics[width=  1 \columnwidth]{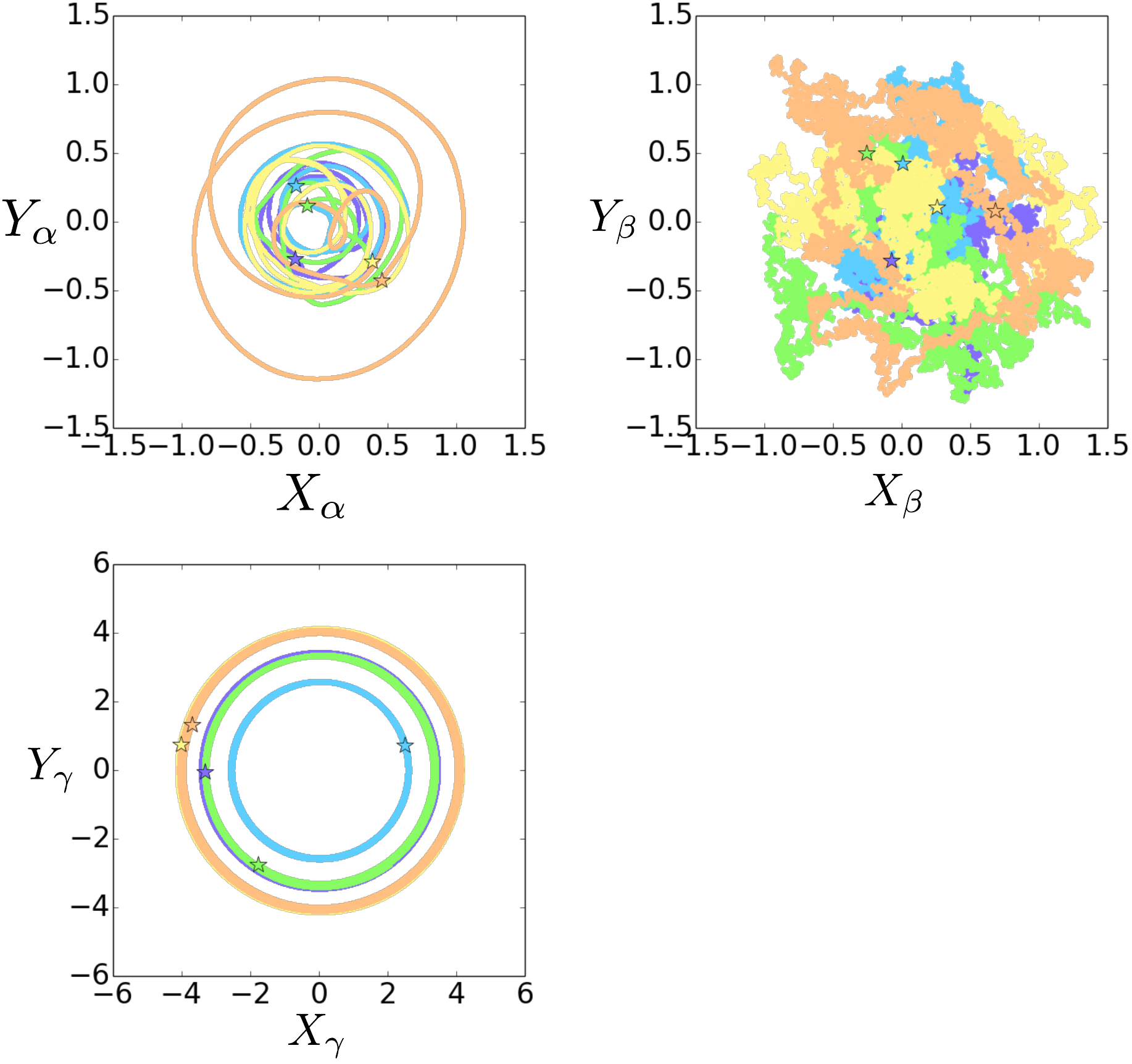} 
\caption{Five different simulations of the classical stochastic equations \eqref{ito}, with $N_a=0$, $N_b=0.5$ and $N_c=0$ (corresponding to the rightmost column of Fig. \ref{limit} in the main text). For each trajectory, only the last 20000 steps (i.e. the asymptotic regime) are shown. Final points are marked by a star. Other parameters are specified in the main text.} \label{examples}
\end{figure}

The differential equations \eqref{stochastic} make sense only with respect to stochastic integration \cite{Gardiner1994b}. In simple words, for simulating the dynamics, we must take finite increments over a small time step $dt$:\\
\footnotesize
\begin{align}
d{\alpha}&\!=\! \left[- i \Delta \alpha  + i g (\alpha + \beta)(\gamma^* + \gamma) - \frac{\kappa_a}{2} \alpha \right] \!dt + \frac{dW^a_x + i dW^a_y}{\sqrt{2}}, \nonumber \\
d{\beta}&\!=\! \left[+ i g (\alpha + \beta)(\gamma^* + \gamma) - \frac{\kappa_b}{2}\beta \right] \!dt + \frac{dW^b_x + i dW^b_y}{\sqrt{2}},   \label{ItoABC}   \\
d{\gamma}& \!=\! \left[- i \omega_c \gamma  + i g |\alpha + \beta|^2  - \frac{\kappa_c}{2} \gamma \right] \!dt + \frac{dW^c_x + i dW^c_y}{\sqrt{2}}. \nonumber
\end{align}\\ 
\normalsize
where now $dW^\nu_x$ and $dW^\nu_y$  ($\nu=a,b,c $) are independent random increments sampled from Gaussian distributions with zero mean and variances equal to $\sqrt{\kappa_\nu N_\nu dt}$. It is convenient to recast Eq.s \eqref{ItoABC} in terms of the real adimensional position and momentum variables $X_\nu$, $Y_\nu$ ($\nu=\alpha,\beta,\gamma $), defined such that $\alpha= (X_\alpha+i Y_\alpha)/\sqrt{2}$, $\beta= (X_\beta+i Y_\beta)/\sqrt{2}$ and $\gamma= (X_\gamma+i Y_\gamma)/\sqrt{2}$. In this way we obtain a system of real stochastic equations which can be efficiently numerically simulated:  \\
\footnotesize
\begin{align}
&d X_\alpha = \left\{ \Delta Y_\alpha - g \sqrt{2} (Y_\alpha+Y_\beta) X_\gamma - \frac{\kappa_a}{2} X_\alpha \right\} dt + dW^a_x,\nonumber\\
&d Y_\alpha = \left\{ -\Delta X_\alpha + g \sqrt{2} (X_\alpha+X_\beta) X_\gamma - \frac{\kappa_a}{2} Y_\alpha\right\} dt + dW^a_y,\nonumber\\
&d X_\beta = \left\{- g \sqrt{2} (Y_\alpha+Y_\beta) X_\gamma - \frac{\kappa_b}{2} X_\beta\right\} dt + dW^b_x,\nonumber\\
&d Y_\beta = \left\{+ g \sqrt{2} (X_\alpha+X_\beta) X_\gamma - \frac{\kappa_b}{2} Y_\beta\right\} dt + dW^b_y,\nonumber\\
&d X_\gamma = \left\{ \omega_c Y_\gamma - \frac{\kappa_c}{2} X_\gamma \right\} dt + dW^c_x,  \label{ito} \\
&d Y_\gamma =\left\{ -\omega_c X_\gamma \!+\! \frac{g}{\sqrt{2}} \left[ (X_\alpha\!+\!X_\beta)^2 \!+\! (Y_\alpha\!+\!Y_\beta)^2\right] \!-\! \frac{\kappa_c}{2} Y_\gamma\right\} dt + dW^c_y,\nonumber
\end{align}\\
\normalsize
We then fix $dt=10^{-3}/\omega_c$ ($1/\omega_c$ being the smallest timescale in the system). Starting with initial conditions $X_\alpha(0)=Y_\alpha(0)=X_\beta(0)=Y_\beta(0)=X_\gamma(0)=Y_\gamma(0)=0$, we add $10^7$ subsequent increments so that the total evolution time becomes $T = 10^7 dt = 10^4/\omega_c \gg 1/\kappa_a,1/\kappa_b,1/\kappa_c$ and the final points are distributed consistently with the stationary state of the system. Collecting $10^4$ different trajectories, we can finally reconstruct the steady-state distribution in phase-space (which is shown in Fig.\ \ref{limit} in the main text) and extract all the desired statistics.\\
\begin{table}[h!]
\begin{tabular}{c|c}step time (dt) & $10^{-3}/\omega_c$ \\\hline number of steps per trajectory & $10^7$ \\\hline number of trajectories & $10^4$\end{tabular}\vspace{10pt}
\caption{Parameters used in the simulation of the classical stochastic equations. Other system parameters are specified in the main text.}
\end{table}\\
In Fig. \ref{examples} we plot five of these classical trajectories, simulated for $N_a=0$, $N_b=0.5$ and $N_c=0$ (this corresponds to the rightmost column of Fig.\ \ref{limit} in the main text, where all other parameters are also specified). To highlight the asymptotic regime, only the last 20000 steps are plotted for each trajectory. Since the frequency of the oscillator $b$ is brought to zero in the rotating frame, the Brownian nature of the motion becomes evident. On the contrary, the oscillator $c$ clearly shows limit cycles of fixed amplitude and random phase.\\

\subsection{Computation of the free energy.}
Through the simulation described above, we collect the ending points of $10^{4}$ trajectories. These points are distributed accordingly to the stationary probability distribution of the $X_{\nu}$, $Y_{\nu}$ variables ($\nu=\alpha,\beta,\gamma $). Therefore, we can reconstruct the free energy of the state, required to characterize the thermodynamic performance of our engine form a resource theory point of view. We want to compute the quantity ${ F = E - k_{B} T H }$ where ${ E = \hbar \omega \left< X_{\gamma}^{2} + Y_{\gamma}^{2} \right>/2 }$ is the mean energy and $H$ is the Shannon entropy of the $X_{\gamma}$, $Y_{\gamma}$ probability distribution. To estimate the latter, we discretize the phase space as a square lattice of step $\Delta$ and take the frequencies $p_{i,j}$ of points appearing in the site $\{i,j\}$ of the grid. Finally, we compute the entropy as ${ H = -\sum_{i,j} p_{i,j} log(p_{i,j}) }$.

We also need the free energy of the Gibbs state at temperature $T$. Although we know the analytic expression, a comparison with the approximate free energy of the limit cycle would not be very meaningful. Instead, we also simulate the Gibbs distribution by putting $g=0$ (no optomechanical coupling) in Eqs.~\eqref{ito} and compute its free energy $F_{G}$ according to the procedure described above. By doing so we remove the problem of the entropy dependence on the step size $\Delta$, since this disappears in the difference $F - F_{G}$ if we use the same step. We have to be careful however in not choosing a very small step (that would give either $0$ or $1$ point in each cell so that $p_{i,j}$ is either $0$ or $1/10^{4}$) or a very big step (that would bring almost all points to a single cell so that a single $p_{i,j}$ is equal to $1$).

\end{document}